\documentclass[twocolumn,showpacs,preprintnumbers,amsmath,amssymb,aps]{revtex4}

\usepackage{graphicx}
\usepackage{dcolumn}
\usepackage{bm}

\begin{document}


\title{Origin of the Inverse Spin Switch Effect in Superconducting Spin Valves}

\author{J. Zhu}
\author{X. Cheng}
\author{C. Boone}
\author{I. N. Krivorotov}
\affiliation{%
Department of Physics and Astronomy, University of California, Irvine, CA 92697-4575
}%




\begin{abstract}

The resistance of a ferromagnet/superconductor/ferromagnet (F/S/F) spin valve near its superconducting transition temperature, $T_c$, depends on the state of magnetization of the F layers. This phenomenon, known as spin switch effect (SSE), manifests itself as a resistance difference between parallel ($R_P$) and antiparallel ($R_{AP}$) configurations of the F layers. Both standard ($R_{P}>R_{AP}$) and inverse ($R_{P}<R_{AP}$) SSE have been observed in different superconducting spin valve systems, but the origin of the inverse SSE was not understood. Here we report observation of a coexistence of the standard and inverse SSE in Ni$_{81}$Fe$_{19}$/Nb/Ni$_{81}$Fe$_{19}$/Ir$_{25}$Mn$_{75}$ spin valves. Our measurements reveal that the inverse SSE arises from a dissipative flow of vortices induced by stray magnetic fields from magnetostatically coupled N\'eel domain wall pairs in the F layers.

\end{abstract}

\maketitle

Multilayers of ferromagnetic and nonmagnetic materials have been a focus of intensive research over the past two decades. Several spin-dependent transport effects such as giant magnetoresistance~\cite{baibich1988,binasch1989} and tunneling magnetoresistance~\cite{moodera1995,miyazaki1995} have been discovered in systems where the nonmagnetic material is a normal metal or an insulator. An equally rich set of magnetotransport phenomena ~\cite{bergeret2005,buzdin2005,villegas2006} including $\pi$-shift in S/F/S Josephson junctions~\cite{kontos2002}, oscillations of $T_c$ with the F layer thickness in F/S bilayers~\cite{zdravkov2006} and spin-triplet Josephson effect~\cite{keizer2006} were recently observed in multilayers of ferromagnetic and superconducting materials.

Despite this recent progress, several phenomena found in F/S multilayers are poorly understood, including the magnetotransport properties of F/S/F spin valve structures. It is well established that the resistance of F/S/F spin valves near $T_c$ depends on the state of magnetization of the F layers~\cite{gu2002,potenza2005,moraru2006a,moraru2006b,miao2008,rusanov2006,steiner2006,stamopoulos2007,singh2007a,pena2005}. However, opposite signs of this effect were observed in different spin valve systems. Several groups found a higher spin valve resistance in the parallel (P) state of the two F layers than in the antiparallel (AP) state ($R_{P}>R_{AP}$)~\cite{gu2002,potenza2005,moraru2006a,moraru2006b,miao2008}. Known as the standard SSE, this phenomenon arises from Cooper pair breaking by the exchange fields of the F layers. Exchange fields induced by two F layers in the S layer partially cancel in the AP configuration, leading to a higher $T_c$ (and lower $R_{AP}$). In the P state, the exchange fields add and thereby more efficiently suppress superconductivity leading to a lower $T_c$ (and higher $R_P$)~\cite{buzdin2005,halterman2007}.

A number of groups have also observed an inverse SSE where the resistance of a superconducting spin valve near $T_c$ rises above the $R_P$ value during the F layer magnetization reversal ~\cite{rusanov2006,singh2007a,steiner2006,stamopoulos2007}. This resistance rise was interpreted as evidence of $R_{P}<R_{AP}$ ~\cite{rusanov2006} because partial AP alignment of the F layers is achieved in the reversal process. Two mechanisms of the inverse SSE have been proposed. One mechanism attributes a lower $T_c$ in the AP configuration to enhanced accumulation of quasiparticles in the S layer resulting in superconducting gap suppression ~\cite{rusanov2006,singh2007a}. The other mechanism is purely magnetostatic. In this mechanism, incomplete saturation of magnetization of the F layers in the reversal process generates a stray magnetic field normal to the S layer and gives rise to vortex flow resistance ~\cite{steiner2006,stamopoulos2007}.

In this Letter, we experimentally determine the origin of the inverse SSE in niobium/permalloy (Py=Ni${}_{81}$Fe${}_{19}$) F/S/F spin valves. We make measurements of SSE in a number of Nb/Py heterostructures that include F/S/F/AF exchange-biased spin valves (AF is antiferromagnetic Ir${}_{25}$Mn${}_{75}$), F/S/F trilayers and F/S bilayers. In exchange-biased spin valves, we observe a coexistence of the standard and inverse SSE in which three resistance states are found: low resistance AP state, intermediate resistance P state and high resistance D state with multiple domains in both F layers. The standard SSE corresponds to the AP $\leftrightarrow$ P transition while the inverse SSE corresponds to the P $\leftrightarrow$ D transition. Observation of coexistence of the standard and inverse SSE in the same sample rules out the quasiparticle accumulation mechanism as the origin of the inverse SSE and lends support to the vortex flow resistance mechanism. Our measurements of the inverse SSE in F/S/F trilayers and F/S bilayers confirm the magnetostatic origin of this effect.

We deposit the samples onto thermally oxidized Si substrates by magnetron sputtering in a high-vacuum system with a base pressure of $5.0\times10^{-9}$ Torr. A magnetic field of 250 Oe is applied in the plane of the samples during growth to induce uniaxial magnetic anisotropy in the Py layers and to set the direction of the exchange bias field. We make three types of Nb/Py heterostructures: Py(6)/Nb(23)/Py(6)/Ir${}_{25}$Mn${}_{75}$(10) exchange-biased spin valves, Py(4)/Nb(25)/Py(4) trilayers and Nb(18)/Py(4) bilayers (thicknesses are in nm). The multilayers are capped with a 4 nm thick Al layer that is not superconducting in the temperature range employed in our measurements. For electrical transport studies, we pattern the samples into 200 $\mu$m-wide Hall bars so that a 20 $\mu$A probe current flows parallel to the easy axis of the growth-induced magnetic anisotropy. Four-point magnetoresistance measurements are made in a continuous flow ${}^4$He cryostat with a temperature stability of $\pm$0.1 mK.

\begin{figure}
\includegraphics[width=.5\textwidth]{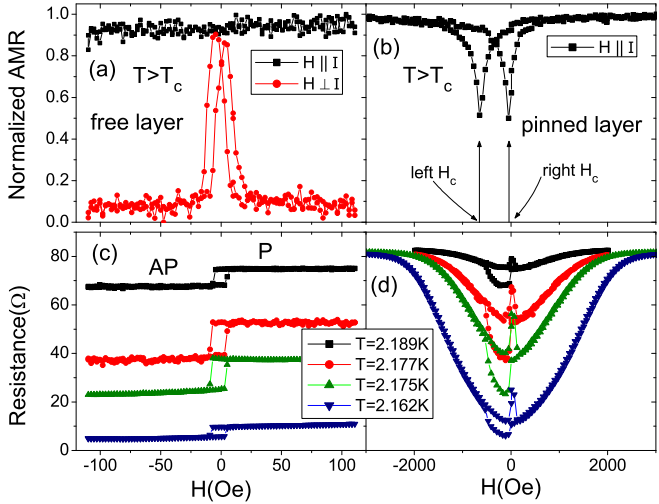}
\caption{\label{fig:biased} (color online). (a) Minor and (b) major normalized AMR hysteresis loops of Py(6)/Nb(23)/Py(6)/Ir${}_{25}$Mn${}_{75}$(10) measured at T = 4.2 K ($T > T_c$). The normalized AMR is defined as $(R-R_0)/\Delta R$. (c) Minor and (d) major easy-axis MR hysteresis loops measured at four temperatures near $T_c$. The minor loop exhibits standard SSE, while the major loop exhibits both standard and inverse SSE.}
\end{figure}

We employ anisotropic magnetoresistance (AMR) to characterize magnetization reversal in Py/Nb heterostructures. Due to AMR, the resistance of a Py film, $R$, depends on  the angle, $\theta$, between the directions of current and magnetization: $R = R_0+\Delta R cos^2(\theta)$. Since $\Delta R > 0$ for Py, higher resistance  corresponds to a greater fraction of magnetization parallel to the current direction ~\cite{gredig2002}. Figures~\ref{fig:biased} (a) and (b) show resistance of a Py(6)/Nb(23)/Py(6)/Ir${}_{25}$Mn${}_{75}$(10) spin valve at $T$ = 4.2 K ($T > T_c$) as a function of magnetic field applied in the plane of the sample parallel (easy axis) and perpendicular (hard axis) to the current direction. Fig.~\ref{fig:biased}(a) displays the easy- and hard-axis AMR minor hysteresis loops of the spin valve where magnetization of the free layer is reversed by a $\pm$ 120 Oe magnetic field while magnetization of the pinned layer remains fixed in the exchange bias field direction. In the hard-axis loop, resistance of the free layer continuously varies between $R_0+\Delta R$ and $R_0$ and saturates for $|H| >$20 Oe. These data show that the reversal process is coherent rotation of magnetization in the uniaxial anisotropy field of approximately 20 Oe. In the easy-axis minor loop, resistance remains nearly constant, consistent with reversal by propagation of a small number of domain walls~\cite{gredig2002}. Fig.~\ref{fig:biased}(b) shows the easy-axis major hysteresis loop where a $\pm$ 3000 Oe magnetic field reverses the magnetizations of both Py layers. This figure demonstrates that the reversal process of the pinned layer is different from that of the free layer. Indeed, at both the left (-650 Oe) and right (-40 Oe) coercive fields ($H_c$) of the pinned layer, magnetization develops a large component perpendicular to the easy axis as evidenced by $R \approx R_0+\Delta R/2$ at the coercive fields.

\begin{figure}[b]
\includegraphics[width=0.5\textwidth]{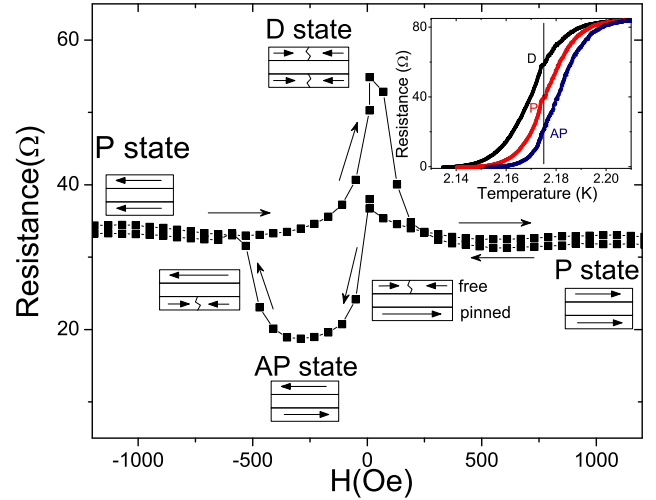}
\caption{\label{fig:RH3states} (color online). Resistance versus field of Py(6)/Nb(23)/Py(6)/Ir${}_{25}$Mn${}_{75}$(10) at $T$ = 2.175 K with a background linear in $|H|$ subtracted from the data. Three distinct resistance states are observed: the AP, the P and the domain state, D. The inset shows resistance versus temperature for the three resistance states.}
\end{figure}

The Py(6)/Nb(23)/Py(6)/Ir${}_{25}$Mn${}_{75}$(10) sample becomes superconducting at $T_c \approx$ 2.175 K (see the inset in Fig.~\ref{fig:RH3states}). Fig.~\ref{fig:biased}(c) and (d) display the minor and major easy-axis magnetoresistance (MR) hysteresis loops of this sample measured at four temperatures near $T_c$. The minor loop in Fig.~\ref{fig:biased}(c) exhibits standard SSE ($R_{P}>R_{AP}$) when the free layer magnetization is switched between P and AP states. However, unusual magnetoresistance behavior is observed in the major loop in Fig.~\ref{fig:biased}(d). When magnetic field is swept from negative to positive field, the resistance of the sample increases well above that of the P state in a narrow field range near zero field. To better illustrate this surprising effect, Fig.~\ref{fig:RH3states} shows the $T$ = 2.175 K data after subtraction of an approximately linear $R(|H|)$ background due to field-induced suppression of $T_c$.

Figure~\ref{fig:RH3states} shows three different resistance states (P, AP and D) of the superconducting spin valve. The P state is initially achieved in a large positive field where both Py layers are saturated along the applied field direction. When the field is swept to a small negative value, magnetization of the free layer reverses and the AP state with $R_{AP}<R_{P}$ is reached. In this reversal process, the pinned layer remains in a single domain state with its magnetization in the exchange bias field direction. Sweeping the field to a large negative value past the left coercive field of the pinned layer reverses magnetization of the pinned layer and the P state is reached again. During this reversal process,  the free layer remains in a single domain state with its magnetization aligned in the external field direction. Sweeping the field from a large negative value to zero results in an increase of resistance near zero field to a value greater than that of the P state. For this field sweep direction, the coercive fields of both Py layers are close to zero and thus multiple domains are present in both Py layers near zero field (D state).

We note that in previous studies of systems exhibiting the inverse SSE~\cite{rusanov2006,steiner2006,stamopoulos2007}, only P and D states could be prepared because the coercive fields of the two F layers were not sufficiently different from each other to achieve the AP state. In previous studies of systems where the standard SSE was observed~\cite{moraru2006a,moraru2006b}, only P and AP states could be prepared because the coercive fields of the F layers were significantly different from each other and the D state with multiple domains in both F layers could not be prepared. The F/S/F/AF spin valve samples studied in this Letter allow us to prepare all three states (P, AP and D)in the same system. The left coercive field of the pinned layer of this sample is much greater than the left coercive field of the free layer, which allows us to achieve the fully aligned AP state. In contrast, the right coercive field of the pinned layer is similar to the right coercive field of the free layer, which allows us to prepare the D state. As we discuss below, the ability to prepare all three states (P, AP and D) in the same system is crucial for understanding the inverse SSE.

Since the multi-domain D state can be viewed as composed of P- and AP-aligned local domains, an assumption was made in Ref.~\cite{rusanov2006} that the resistance of the D state, $R_D$, was a weighted average of resistances of the P and AP states. If this assumption is correct then experimental observation of the inverse SSE, $R_{D}>R_{P}$, proves that $R_{AP}>R_{P}$. Our exchange-biased spin valves allow us to test this assumption directly since all three states (P, AP and D) can be prepared in this system. The experimentally found relation between resistances of the P, AP and D states, $R_{D}>R_{P}>R_{AP}$, shown in Fig.~\ref{fig:RH3states} is inconsistent with the assumption that $R_D$ is a weighted average of $R_P$ and $R_{AP}$. Therefore, observation of the inverse SSE, $R_{D}>R_{P}$, does not prove that $R_{AP}>R_{P}$. In fact, the opposite ($R_{AP}<R_{P}$) is true as evidenced by the data in Fig.~\ref{fig:RH3states}. Therefore, the observed resistance increase in the D state cannot result from the AP alignment of the F layers, and the mechanism of quasiparticle accumulation in the AP state of a superconducting spin valve proposed in Ref.~\cite{rusanov2006} cannot be the origin of the inverse SSE. Fig.~\ref{fig:RH3states} also shows that a multi-domain state in only one of the F layers does not give rise to the inverse SSE -- both F layers must be in multi-domain states for a significant inverse SSE to be observed.

The D state can affect the resistance of a superconducting spin valve near $T_c$ via two mechanisms. One mechanism is domain-wall superconductivity ~\cite{yang2004,rusanov2004,zhu2008} where the pair-breaking exchange field induced by an F layer in the S layer is reduced near domain walls, leading to a decrease of resistance in the D state. Another mechanism is dissipation by vortices induced in the S layer by the stray magnetic field from domain walls in the F layers, leading to an increase of resistance in the D state  ~\cite{stamopoulos2007}. The data in Fig.~\ref{fig:RH3states} demonstrate that the vortex flow resistance dominates the domain-wall superconductivity in our F/S/F/AF samples.

\begin{figure}
\includegraphics[width=0.4\textwidth]{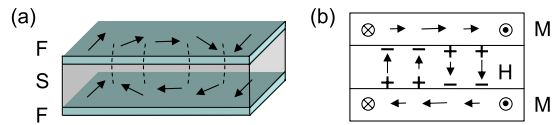}
\caption{\label{fig:wwcoupling} (color online). (a) Schematic of a pair of magnetostatically coupled N\'eel domain walls in a F/S/F trilayer and (b) Side view of magnetization, $M$, in the domain walls and stray field, $H$, generated by the domains walls in the spacer.}
\end{figure}

We estimate the magnitude of resistance increase induced by a stray field from domain walls in the D state.  Slonczewski and Middelhoek~\cite{slonczewski1966, slonczewski1965, middelhoek1966} demonstrated that positions of N\'eel domain walls in two F layers separated by a nonmagnetic spacer are correlated -- the wall in the top F layer is located directly above the wall in the bottom F layer, as shown in Fig.~\ref{fig:wwcoupling}. This correlated domain wall state decreases magnetostatic energy of the system via magnetic flux line closure through the spacer, and increases the normal component of the average stray field in the spacer. The correlated domain walls induced by magnetostatic coupling were previously observed in magnetic tunnel junctions~\cite{gider1998} and giant magnetoresistance spin valves~\cite{lew2003}. The average magnitude of the stray magnetic field in the spacer between a domain wall pair, $\overline{|H_s|}=8\pi D M_s/a$, is calculated from the magnetic scalar potential given by Eq.(4) of Ref.~\cite{slonczewski1965}, where $D$ = 6 nm is the thickness of the Py layer, $M_s$ = 800 emu/cm$^3$ is the saturation magnetization of Py and $a$ is the domain wall width. The domain wall width of the N\'eel wall pairs is $a=\pi \sqrt{2[A+\pi M_s^2 D (b/2+D/3)]/K}$~\cite{middelhoek1966}, where $A=10^{-6}$ erg/cm is the exchange constant, $b$ = 23 nm is the thickness of the spacer and $K$ is the uniaxial anisotropy constant of the F layers. This equation for $a$ was derived assuming that both F layers have the same uniaxial anisotropy. This assumption is not valid for our system where the uniaxial anisotropy of the pinned layer is strongly enhanced by exchange bias~\cite{krivorotov2003}. However, bounds on $a$ can be calculated using the uniaxial anisotropy constants of the free and pinned layers. The anisotropy constant of the free layer, $K=8\times10^3$ erg/cm$^3$, is given by the free layer anisotropy field $H_a\approx$ 20 Oe. For the pinned layer, we use the coercivity ($H_c\approx$ 300 Oe) as an estimate of $H_a$, which gives $K=1.2\times10^5$ erg/cm$^3$. These values of $K$ give bounds on the domain wall width, 0.2 $\mu$m $< a <$ 0.8 $\mu$m, and the average normal stray field between a domain wall pair, 150 Oe $<\overline{|H_s|}<$ 600 Oe. We also directly measure vortex flow resistance of the spin valve in a uniform external magnetic field applied perpendicular to the sample plane at $T$ = 2.175 K and find that a field of 20 Oe is sufficient to increase resistance of the sample from $R_{P}$ to $R_{D}$. Comparing this field to $\overline{|H_s|}$, we estimate that domain walls occupy 3\% to 13\% of the spin valve area in the D state, corresponding to an average domain width $\approx$ 12 $\mu$m typical for exchange-biased Py films~\cite{gillies1995}.

\begin{figure}
\includegraphics[width=0.5\textwidth]{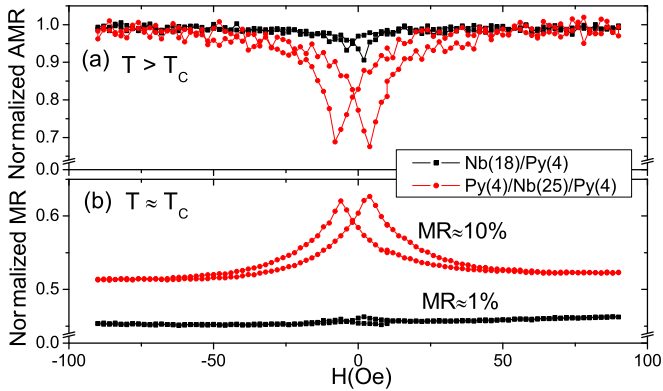}
\caption{\label{fig:RHbitri} (color online). (a) Normalized easy-axis AMR at T = 4.2 K ($T>T_c$) and (b) Normalized easy-axis MR at $T \approx T_c$ for Nb(18)/Py(4) bilayer and Py(4)/Nb(25)/Py(4) trilayer.}
\end{figure}

Further evidence of magnetostatic origin of the inverse SSE is given by MR measurements of Nb(18)/Py(4) bilayers and Py(4)/Nb(25)/Py(4) trilayers. Fig.~\ref{fig:RHbitri}(a) shows easy-axis AMR hysteresis loops of the bilayer and trilayer measured at $T > T_c$. These measurements demonstrate that magnetization reversal processes of bilayers and trilayers are significantly different. Small deviations of resistance of the bilayer from $R_0+\Delta R$ in the reversal process show that only a small fraction of magnetization is perpendicular to the easy axis, consistent with reversal by propagation of a small number of domain walls. In contrast, the resistance of the trilayer decreases to $\approx R_0+2/3 \Delta R$ at the coercive fields, consistent with formation of a large number of domain walls in the reversal process. Using resistance minima of the AMR hysteresis loops in Fig.~\ref{fig:RHbitri}(a) as a measure of the domain wall areal density, we estimate that four times as many domain walls are formed in the trilayer as in the bilayer during the reversal process. Fig.~\ref{fig:RHbitri} (b) shows easy-axis MR of the bilayer and the trilayer at $T$ near $T_c$ normalized to resistance at $T>T_c$. Both the trilayer and the bilayer exhibit inverse SSE but the magnitude of this effect is significantly larger in the trilayer (10\%) than in the bilayer (1\%). Given that stray field from the domain wall pairs in the trilayer is approximately twice that from single walls in the bilayer, and the density of domain walls in the trilayer is approximately four times that in the bilayer, the magnitude of the inverse SSE is expected to be a factor of eight greater in the trilayer than in the bilayer. This estimate is consistent with the data in Fig.~\ref{fig:RHbitri}(b).

In conclusion, we observe a coexistence of standard and inverse spin switch effects in Py/Nb/Py/Ir${}_{25}$Mn${}_{75}$ exchange-biased spin valves. This observation allows us to determine that the inverse spin switch effect originates from vortex flow resistance induced by stray magnetic fields from domain walls in the Py layers. A large magnitude of the inverse spin switch effect observed in Py/Nb/Py/Ir${}_{25}$Mn${}_{75}$ and Py/Nb/Py spin valves compared to Py/Nb bilayers is consistent with formation of magnetostatically coupled, spatially correlated N\'eel domain wall pairs in the pinned and free Py layers of the spin valves. This magnetostatic coupling increases both the domain wall density and the magnitude of the domain wall stray fields, leading to the inverse spin switch effect enhancement in spin valve samples. Our work demonstrates that the inverse spin switch effect in superconducting spin valves is magnetostatic in origin and is not caused by quasiparticle accumulation in the superconducting layer in the antiparallel state of the spin valve.

\end{document}